# Hidden peculiar magnetic anisotropy at the interface in a ferromagnetic perovskite-oxide heterostructure

Le Duc Anh,[1,2,*] Noboru Okamoto,[1] Munetoshi Seki,[1] Hitoshi Tabata,[1,3] Masaaki Tanaka,[1,3,**] and Shinobu Ohya[1,2,3,***]

[1]*Department of Electrical Engineering and Information Systems, The University of Tokyo, 7-3-1 Hongo, Bunkyo-ku, Tokyo 113-8656, Japan*
[2]*Institute of Engineering Innovation, Graduate School of Engineering, The University of Tokyo, 7-3-1 Hongo, Bunkyo-ku, Tokyo 113-8656, Japan*
[3]*Center for Spintronics Research Network (CSRN), The University of Tokyo, 7-3-1 Hongo, Bunkyo-ku, Tokyo 113-8656, Japan*

* Corresponding author: anh@cryst.t.u-tokyo.ac.jp
** Corresponding author: masaaki@ee.t.u-tokyo.ac.jp
*** Corresponding author: ohya@cryst.t.u-tokyo.ac.jp

**Understanding and controlling the interfacial magnetic properties of ferromagnetic thin films are crucial for spintronic device applications. However, using conventional magnetometry, it is difficult to detect them separately from the bulk properties. Here, by utilizing tunneling anisotropic magnetoresistance in a single-barrier heterostructure composed of $La_{0.6}Sr_{0.4}MnO_3$ (LSMO)/ $LaAlO_3$ (LAO)/ Nb-doped $SrTiO_3$ (001), we reveal the presence of a peculiar strong two-fold magnetic anisotropy (MA) along the $[110]_c$ direction at the LSMO/LAO interface, which is not observed in bulk LSMO. This MA shows unknown behavior that the easy magnetization axis rotates by 90° at an energy of 0.2 eV below the Fermi level in LSMO. We attribute this phenomenon to the transition between the $e_g$ and $t_{2g}$ bands at the LSMO interface. Our finding and approach to understanding the energy dependence of the MA demonstrate a new possibility of efficient control of the interfacial magnetic properties by controlling the band structures of oxide heterostructures.**

Control of magnetic anisotropy (MA) is crucial for low-power magnetization reversal in magnetic thin films. From the perspectives of energy efficiency and scalability, gate-voltage control of the MA via modulation of the carrier density and thus, the Fermi level, is highly desirable[1–4]. For efficient control of MA and for developing materials that are suitable for the MA control, it is necessary to understand the MA of the magnetic thin films over a wide energy range; however, there are few studies from this point of view. In ferromagnetic (FM) materials, the MA energy is related to the magnetization-direction dependence of the density of states (DOS) via the spin orbit interaction[5]. Tunneling anisotropic magnetoresistance (TAMR) is a phenomenon observed in tunnel diodes composed of ferromagnetic (FM) layer/ tunnel barrier/ nonmagnetic (NM) electrode. TAMR is defined as the change of the tunnel resistance or conductance d$I$/d$V$, which is proportional to the DOS of the electrodes, when rotating the magnetization of the FM layer[5–10]. Thus, TAMR is useful to understand the magnetic-field direction dependence of the DOS. By measuring TAMR at various bias voltages, one can obtain a high-resolution carrier-energy-resolved map of MA of the FM layer[8–10].

An equally important aspect of TAMR is that it reflects the DOS at the *tunneling interface* of the FM layer, and thus it provides a sensitive probe of the interfacial magnetic properties. Thin film interfaces present both problems and opportunities for exploring new functional devices. As a good example, the "dead layer" at the interface of the perovskite oxide La$_{1-x}$Sr$_x$MnO$_3$ (LSMO, the Sr content $x = 0.3$–0.4), which is one of the most promising materials due to its intriguing magnetic and electrical properties such as the colossal magnetoresistance[11,12], half-metallic band structure[13,14], and high Curie temperature ($T_C$ ~370 K)[11], is a serious problem for its device applications. For the formation of the dead layer, various possible origins have been proposed, such as intermixing of atoms[15], oxygen vacancies[16], lattice distortion[17-19], and $MnO_6$ oxygen octahedral rotations (OOR)[20-22], which induce orbital, charge, and spin reconstruction at the interfaces of LSMO. These studies on dead layers, however, suggest new ways for controlling the interfacial properties at an atomic level, which are not available in the bulk. To this end, the characterization of the interfacial magnetic properties is highly demanded, but it is difficult with conventional magnetometry because the interfacial properties are usually concealed by the dominant signals from the bulk. Here, by utilizing TAMR in an LSMO/LAO/Nb:STO junction, we obtain the carrier-energy dependence of MA of LSMO for the first time. We also reveal a peculiar strong two-fold symmetry component of MA at the LSMO/LAO interface, which is not observed in bulk LSMO. Moreover, this interfacial MA shows unknown behavior that the symmetry axis of this interface MA rotates by 90° at an energy of 0.2 eV below the Fermi level in LSMO. We attribute this phenomenon to the transition between the $e_g$ and $t_{2g}$ bands at the LSMO interface. Our results suggest that controlling the band structure at interfaces will pave a new way for efficient control of the magnetization of FM thin films, which is essential for devices with low-power consumption.

**RESULTS**

**Sample preparation and characterizations**

The heterostructure used in this study consists of LSMO (40 unit cell (u.c.) = 15.6 nm)/ $LaAlO_3$ (LAO, 4 u.c. = 1.6 nm) grown on a $TiO_2$-terminated Nb-doped $SrTiO_3$ (001) substrate (Nb:STO, Nb 0.5% wt.) by molecular beam epitaxy (MBE) (see Fig. 1a and Methods)[23,24]. The *in-situ* reflection high-energy electron diffraction (RHEED) patterns in the [100] direction of the 4-u.c. LAO and 40-u.c. LSMO layers show streaky patterns, and especially LSMO exhibits a bright pattern (Fig. 1b), indicating that the sample surface is atomically flat. In fact, the atomic force microscopy measurements show flat terraces and atomic steps with a height of ~ 0.4 nm, which is equal to one pseudocubic u.c. (Fig. 1c). In the x-ray reciprocal lattice map of the sample measured around the $(204)_c$ and $(\overline{2}04)_c$ reflections of the Nb:STO substrate at room temperature, we see two weaker peaks corresponding to the $(260)_o$ and $(620)_o$ reflections of the LSMO epilayer (we use the subscripts c and o for the pseudocubic and the orthorhombic crystal structures, respectively) (Fig. 1d). These results confirm that the LSMO layer is coherently grown with respect to the Nb:STO substrate. The $(260)_o$ and $(620)_o$ peaks of LSMO have nearly the same out-of-plane reciprocal lattice vector $Q_\perp$, indicating that the $(260)_o$ and $(620)_o$ atomic plane spacings are equal. This is consistent with the common reports on LSMO thin films grown under tensile strain, indicating that the strain effect in LSMO is accommodated equally between the $[100]_c$ and $[010]_c$ directions[22].

For tunneling transport measurements, $600 \times 700$ $\mu m^2$ mesas were formed by standard photolithography and Ar ion milling. The bias polarity is defined so that the current flows from the LSMO layer to the Nb:STO substrate in the positive bias.

**Magnetic anisotropy components of the LSMO/LAO interface**

Figure 2a shows the conduction band (CB) profiles of the LSMO/LAO/Nb:STO tunnel diode under positive and negative bias voltages $V$. The Fermi level $E_F$ is located at $10 - 20$ meV above the CB bottom of STO due to the Nb doping (0.5% wt., the electron density $n = 1 \times 10^{20}$ $cm^{-3}$)[25], while $E_F$ lies in the CB formed by the Mn $3d$-$e_g$ states in LSMO. The LAO layer serves as a tunnel barrier with a height of ~2.4 eV for STO and ~ 2 eV for LSMO[26]. TAMR measurements were conducted as follows: d$I$/d$V$–$V$ curves were measured at 4 K while applying a strong external magnetic field of 1 T, which aligned the magnetization direction parallel to the magnetic field, in various in-plane directions with an angle step of 10°. The change in d$I$/d$V$ when rotating the external magnetic field is attributed to the change in the DOS at the LSMO/LAO interface or the LAO/STO interface. As illustrated in Fig. 2a, at positive (negative) $V$, electrons tunnel from Nb:STO to LSMO (from LSMO to Nb:STO), and thus d$I$/d$V$ probes the DOS of unoccupied (occupied) states in LSMO. We define $\Phi$ as the angle of the magnetization direction from the [100]$_c$ axis in the counter-clockwise direction in the film plane. At each fixed bias $V$ and angle $\Phi$, we calculate $\Delta\left(\dfrac{\mathrm{d}I}{\mathrm{d}V}\right)$ as $\left(\dfrac{\mathrm{d}I}{\mathrm{d}V}-\left\langle\dfrac{\mathrm{d}I}{\mathrm{d}V}\right\rangle_\Phi\right)\Big/\left\langle\dfrac{\mathrm{d}I}{\mathrm{d}V}\right\rangle_\Phi\times100$ (%), where $\left\langle\dfrac{\mathrm{d}I}{\mathrm{d}V}\right\rangle_\Phi$ is defined as the averaged $\dfrac{\mathrm{d}I}{\mathrm{d}V}$ over $\Phi$ at that specific $V$. As seen in the polar plots of $\Delta\left(\dfrac{\mathrm{d}I}{\mathrm{d}V}\right)$ as a function of $\Phi$ and bias $V$ in Fig. 2b, the magnetization direction dependence of the DOS has a mainly two-fold symmetry, but the symmetry axes are different depending on $V$: At $V = -0.1$ V (left panel), the maximum is located at ~ 150°, between the [010]$_c$ and [$\bar{1}$00]$_c$ axes, while it is at 45° (the [110]$_c$ axis) when $V = -0.35$ V (right panel). The whole picture of this behavior for all $V$ is represented in Fig. 2c, where $\Delta$(d$I$/d$V$) is plotted as a function of $\Phi$ at $V$ ranging from –0.5 to 0.5 V. This plot shows a peculiar behavior that the symmetry axis changes at $V$ ~ –0.2 V. We fit the data at each bias $V$ using the following equation:

$$\Delta\left(\frac{\mathrm{d}I}{\mathrm{d}V}\right) = C_{4\langle110\rangle}\cos\left[4\left(\Phi-\frac{\pi}{4}\right)\right] + C_{2[100]}\cos2\Phi + C_{2[110]}\cos\left[2\left(\Phi-\frac{\pi}{4}\right)\right] \qquad (1)$$

Here $C_{4\langle110\rangle}$ is the four-fold component with the symmetry axis of <110>$_c$, $C_{2[100]}$ and $C_{2[110]}$ are the two-fold components with the [100]$_c$ and [110]$_c$ axes, respectively. As shown in Fig. 2b, the fitting curves (red curves) well reproduce the experimental data (blue points). The three anisotropy components estimated in the whole range of $V$ are summarized in Fig. 2d. The two-fold symmetry ($C_{2[100]}$ or $C_{2[110]}$) is stronger than the four-fold symmetry ($C_{4<110>}$) in almost all the bias region. The $C_{4<110>}$ and $C_{2[100]}$ components show a similar $V$-dependence, stretching from –0.45 V to 0.4 V. The sign of $C_{2[110]}$ changes at $V = \sim$–0.2 V, which indicates an opposite dependence of the DOS on the magnetization direction between the regions of $V >$ –0.2 and $V <$ –0.2. Because the change in the DOS when changing the magnetization direction, $i.e.$ $\Delta(\mathrm{d}I/\mathrm{d}V)$, is proportional to the MA energy[5], the sign change of $C_{2[110]}$ indicates a 90°-shift of the magnetization direction where the MA energy becomes minimum, and consequently indicates a 90°-shift of the easy magnetization axis of the $C_{2[110]}$ component from the $[110]_c$ to the $[\bar{1}10]_c$ directions at $V \sim$ –0.2 V when decreasing $V$. This is the main origin of the symmetry axis rotation observed in Figs. 2b and 2c.

## DISCUSSIONS

We discuss the origins of these anisotropy components. The biaxial (four-fold) MA along the $<110>_c$ axes and the uniaxial (two-fold) MA along the $[100]_c$ axis have been reported for LSMO films grown on STO (001) substrates[22,27-29], and thus they are bulk-like properties. The biaxial component originates from the in-plane cubic symmetry of LSMO thin films grown on STO. The uniaxial MA along the $[100]_c$ axis has been attributed to various origins such as step edges[27,28] or different OOR between the $[100]_c$ and $[010]_c$ directions[22,29]. In our study, this uniaxial MA along the $[100]_c$ axis may also be partially contributed from the high-mobility two dimensional electron gas possibly formed at the LAO/STO interface, where the magnetoconductance has been reported to possess the same two-fold symmetry[30]. On the other hand, the uniaxial MA component along the $[110]_c$ direction has never been reported for LSMO. Because the crystal structure of LSMO grown on a substrate with a cubic symmetry such as STO is equivalent between the $[110]_c$ and $[\bar{1}10]_c$ directions, the uniaxial MA along the $[110]_c$ direction is not a bulk property, and thus must be attributed to the LSMO/LAO interface.

To confirm the interface origin of the uniaxial MA component along the $[110]_c$ axis, we measured the planar Hall resistance (PHR) of a Hall bar with a size of $50 \times 200$ μm formed along the $[100]_c$ direction of a reference sample composed of LSMO (40 u.c.)/ LAO (4 u.c.). This sample was grown on a non-doped STO (001) substrate under the same conditions as those for the sample used for the TAMR measurements. PHR is proportional to $\sin\Phi\cos\Phi$, where $\Phi$ is the angle between the magnetization and the current direction[31]. In Fig. 3, we show the PHR measured for the reference sample at various magnetic field directions, where $\theta$ is the angle between the magnetic field and the current flown in the $[100]_c$ axis. $\Delta R$ is defined as the PHR with respect to the one at a zero magnetic field. In contrast to the TAMR results, the PHRs of the reference sample measured when the magnetic field is parallel to the $[110]_c$ and $[\bar{1}10]_c$ directions are identical (*i.e.* four-fold like) except the opposite signs, showing *no* clue of the uniaxial MA along the $[110]_c$ axis. On the other hand, the PHRs clearly show a dominant uniaxial MA with the easy axis along the $[100]_c$ direction. Because PHR mainly reflects the bulk properties of LSMO, our results confirm that the $C_{4<110>}$ and $C_{2[100]}$ components are the MA inherited from bulk LSMO, while the $C_{2[110]}$ component is the MA that appears only at the LSMO/LAO interface.

Because the $C_{2[110]}$ component is attributed to a symmetry breaking between the $[110]_c$ and $[\bar{1}10]_c$ directions occurring locally near the LSMO/LAO interface, the OOR mechanism is most likely the origin of $C_{2[110]}$. Recently, it has been clarified that adjacent corner-sharing oxygen octahedra in LAO grown on STO (001) rotate in the opposite directions around the $[11\bar{1}]_c$ axis[32,33], and that the OOR in the underlayer is transferred to the first 3–4 u.c. layers of LSMO[20,21](Fig. 4a). Figure 4b shows four adjacent $MnO_6$ octahedra in a $(001)_c$ plane of LSMO near the LSMO/LAO interface. We see that under the OOR around the $[11\bar{1}]_c$ axis illustrated in Fig. 4a, the oxygen octahedra located along the $[110]_c$ direction rotate in the same direction (see the green oxygen spheres around Mn1 and Mn3). This rotation direction is the opposite to that of the adjacent rows of Mn atoms (see Mn2 and Mn4). When Fig. 4b is projected in the $(\bar{1}10)_c$ plane as shown in Fig. 4c, one can see that the vertical O–Mn1–O and O–Mn3–O bonds (see the green spheres) remain nearly perpendicular to the $(001)_c$ plane (dotted line). Because the vertical O–Mn1–O bonds are rotated in the same direction as the O–Mn3–O bonds, the hopping integral ($t_{13}$) between Mn1 and Mn3 remains nearly unchanged from the value of the bulk. On the other hand, when Fig. 4b is projected in the $(110)_c$ plane (Fig. 4d), one can see that the $MnO_6$ oxygen octahedra around Mn2 and Mn4 are largely tilted to the left. This decreases the hopping integral ($t_{24}$) between Mn2 and Mn4. The difference between $t_{13}$ and $t_{24}$ yields the anisotropic DOS between the $[110]_c$ and $[\bar{1}10]_c$ directions and consequently induces $C_{2[110]}$ at the LSMO/LAO interface.

The most striking feature found in our study is the sign reversal of $C_{2[110]}$ at $V = -$

0.2 V (Fig. 2d). This behavior is likely related to the band structure of LSMO, as explained below. In Fig. 2d, one can see that $C_{4<110>}$ and $C_{2[100]}$ are dominant at $V = 0$ and show similar $V$ dependence in all the $V$ region. These results indicate that both originate from the same band located around $E_F$ of LSMO, *i.e.* the up-spin Mn $3d$-$e_g$ band. The $C_{4<110>}$ and $C_{2[100]}$ components disappear at ~ $V = -0.45$ V, which means that $E_F$ is located at ~0.45 eV above the bottom of the $e_g$ band. This is consistent with the results of angle-resolved photoemission spectroscopy (ARPES) measurements for LSMO[34]. Therefore, the emergence of positive $C_{2[110]}$ below $V = -0.2$ V is likely associated with the $t_{2g}$ band, which is located below the $e_g$ band. Although the $t_{2g}$ state is located at 0.5–1 eV below $E_F$ in bulk LSMO[34], it is thought to be largely pushed up by the polar mismatch at the LSMO/LAO interface[35]. Thus, we attribute the sign change of $C_{2[110]}$ to the transition from the $e_g$ band ($V > -0.2$ V) to the $t_{2g}$ band ($V < -0.2$ V) at the LSMO interface. As mentioned above, due to the OOR at the LSMO/LAO interface, the DOSs of both the $e_g$ and $t_{2g}$ bands in the $[110]_c$ direction are larger than those in the $[\bar{1}10]_c$ direction. However, the relationship between the DOS and the magnetization in these two band components is opposite: It is known that the electron transfer via the $e_g$ orbitals enhances the double exchange interaction and strengthens the ferromagnetism, while the one between the $t_{2g}$ orbitals enhances the super-exchange interaction and weakens the ferromagnetism[36]. Therefore, the enhancement of DOS in the $[110]_c$ direction relative to that in the $[\bar{1}10]_c$ direction makes the $[110]_c$ axis the easy magnetization direction in the case of the $e_g$ orbitals, while hard magnetization direction in the case of the $t_{2g}$ orbitals. A transition between these two bands at $V \approx -0.2$ V thus leads to a change in the MA energy corresponding to the $C_{2[110]}$ component from minimum to maximum along the $[110]_c$ axis. This is consequently observed by TAMR as the opposite magnetization direction-dependence of the DOS along the $[110]_c$ axis[5].

In summary, using TAMR measurements, we have successfully obtained a high-resolution map of the MA spectrum of LSMO for the first time. In addition to the biaxial MA along $\langle 100 \rangle_c$ and the uniaxial MA along $[100]_c$, which originate from bulk LSMO, we found a peculiar uniaxial MA along the $[110]_c$, which is attributed to the LSMO/LAO interface. The symmetry axis of this interface MA rotates by 90° at an energy of 0.2 eV below $E_F$ of LSMO, which is attributed to the transition from the $e_g$ band (>–0.2 eV) to the $t_{2g}$ band (<–0.2 eV). These findings hint an efficient way to control the magnetization at the LSMO thin film interfaces, as well as confirm the rich of hidden properties at thin film interfaces that can be revealed only by interface-sensitive probes. This work also suggests the use of TAMR measurement as a simple but highly sensitive method for characterizing interfacial magnetic properties of magnetic tunnel junctions, which is important for developing spintronic devices.

## METHODS

The heterostructure used in this study consists of LSMO (40 unit cell (u.c.) = 15.6 nm)/ LaAlO₃ (LAO, 4 u.c. = 1.6 nm) grown on a TiO₂-terminated Nb-doped SrTiO₃ (001) substrate (Nb:STO, Nb 0.5% wt.) by molecular beam epitaxy (MBE) with a shuttered growth technique[23,24]. The fluxes of La, Sr, Mn, and Al were supplied by Knudsen cells. The LAO and LSMO layers were grown at 730°C with a background pressure of $2\times10^{-4}$ Pa of a mixture of oxygen (80%) and ozone (20%). After the growth, the sample was further annealed at 600°C in ambient atmosphere for 1 hour to reduce the density of oxygen vacancies.

For tunneling transport measurements, a 50-nm-thick Au film was deposited on top of the sample, and $600 \times 700~\mu m^2$ mesas were then formed by standard photolithography and Ar ion milling. Au wires were bonded to the Au electrode and the backside of the Nb:STO substrate by indium.

The $dI/dV$-$V$ and $d^2I/dV^2$-$V$ characteristics were numerically obtained from the $I$-$V$ data with a differential interval of 10 mV. See Supplementary Information for more details on how to extract the $\Delta \left( \dfrac{dI}{dV} \right)$ data plotted in Fig. 2.

**Data Availability**

The datasets generated during and/or analysed during the current study are available from the corresponding author on reasonable request.

**Acknowledgements**

This work was partly supported by Grants-in-Aid for Scientific Research (No. 26249039, No. 23000010) and Project for Developing Innovation Systems of MEXT, Spintronics Research Network of Japan (Spin-RNJ), and the Cooperative Research Project Program of RIEC, Tohoku University. We thank K. Takeshima and T. Matou for technical helps in sample growth.

**Author contributions**

Experiment design and data analysis: L. D. A, S. O.; device fabrication and measurements: L. D. A, N. O, M. S, writing and project planning: L. D. A, S. O., M. T. and H. T. All authors extensively discussed the results and the manuscript.

**Competing financial interests**

The authors declare no competing financial interests.

**Figures and figure legends**

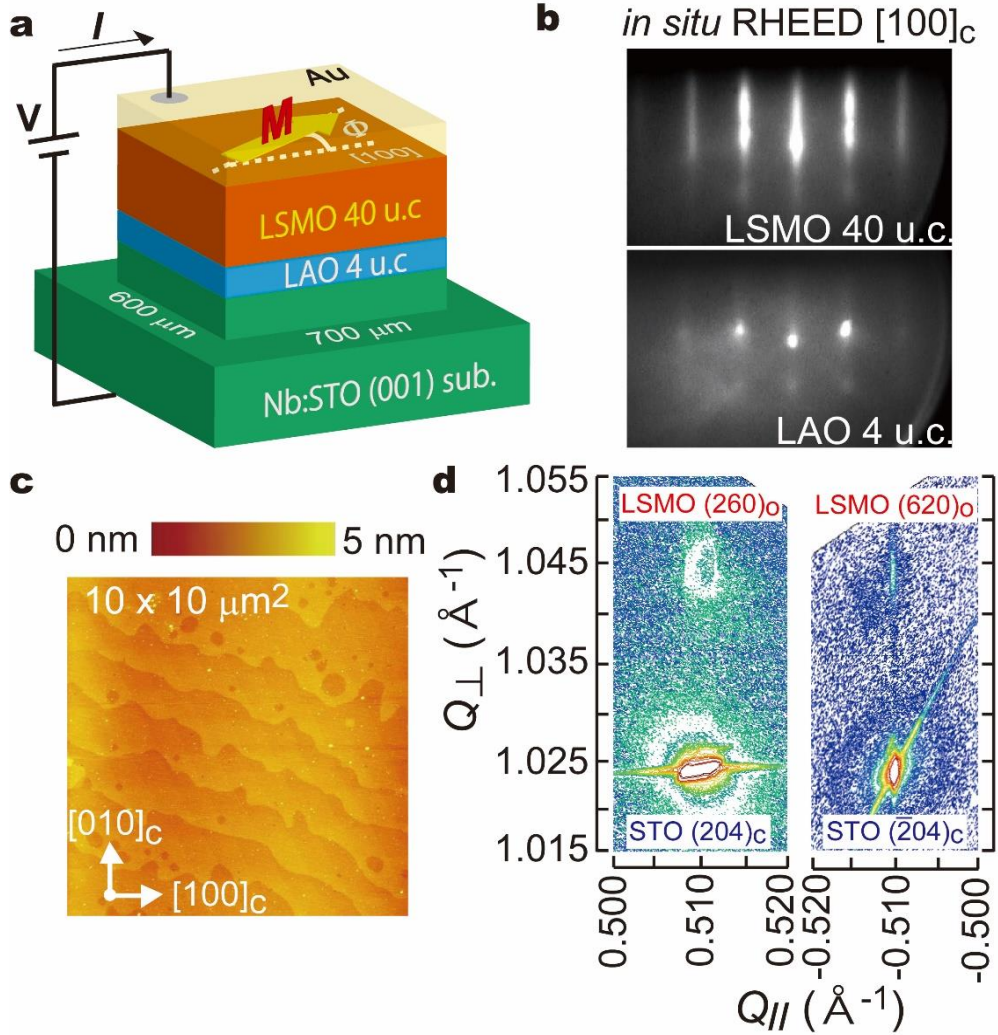

**Figure 1. Sample preparation and characterization.** (**a**) Device structure and tunneling transport measurement configuration of the LSMO/LAO/Nb:STO tunneling diode structure used in this study. (**b**) *In-situ* reflection high-energy electron-diffraction patterns in the $[100]_c$ direction of the LSMO and LAO layers. (**c**) Surface morphology of the LSMO/ LAO/ Nb:STO sample measured by atomic force microscopy. (**d**) Reciprocal lattice maps of the sample measured at room temperature. Here, $Q_{//}$ and $Q_{\perp}$ are the components of the reciprocal lattice vector in the in-plane $[100]_c$ and out-of-plane $[001]_c$ directions, respectively.

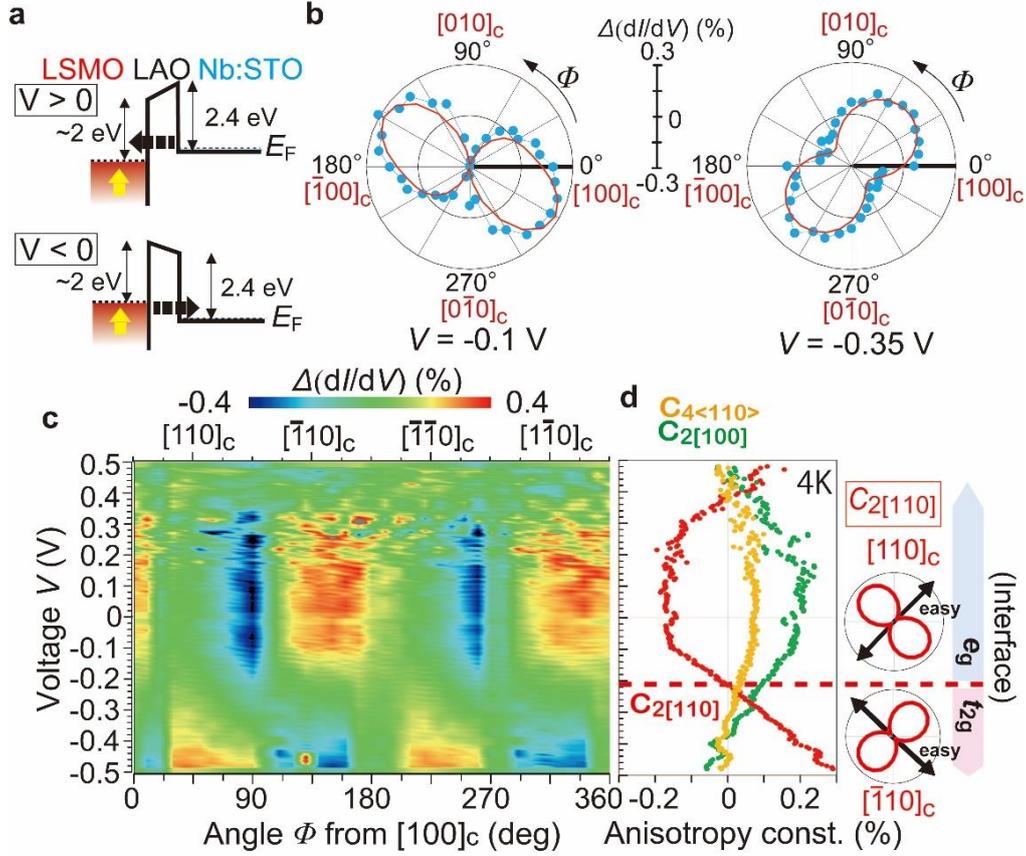

**Figure 2**. **Tunneling anisotropic magnetoresistance results.** (**a**) Conduction band (CB) profiles of the LSMO/ LAO/ Nb:STO tunneling diode under positive and negative bias voltages *V*. The solid and dotted lines represent the top of the CB and the Fermi level $E_F$. At positive (negative) *V*, the electrons tunnel from Nb:STO to LSMO (from LSMO to Nb:STO). (**b**) Polar plots of $\Delta\left(\dfrac{\mathrm{d}I}{\mathrm{d}V}\right) = \left(\dfrac{\mathrm{d}I}{\mathrm{d}V} - \left\langle\dfrac{\mathrm{d}I}{\mathrm{d}V}\right\rangle_\Phi\right) / \left\langle\dfrac{\mathrm{d}I}{\mathrm{d}V}\right\rangle_\Phi \times 100$ (%) as a function of $\Phi$ at *V* = –0.1 and –0.35 V (blue points). Here, $\Phi$ is the magnetic-field angle from the [100]$_c$ axis in the counter-clockwise direction in the film plane, and $\left\langle\dfrac{\mathrm{d}I}{\mathrm{d}V}\right\rangle_\Phi$ is defined as averaged $\dfrac{\mathrm{d}I}{\mathrm{d}V}$ over $\Phi$ at each *V*. The red curves are fitting curves. (**c**) Color plots of $\Delta\left(\dfrac{\mathrm{d}I}{\mathrm{d}V}\right)$ as a function of $\Phi$ and *V*. (**d**) *V*-dependence of the symmetry components $C_{4[110]}$, $C_{2[100]}$, and $C_{2[110]}$. The sign of $C_{2[110]}$ component changes at *V* = –0.2 V, which corresponds to a 90° rotation of the easy magnetization axis of this component. This is attributed to the transition between the $e_g$ band and the $t_{2g}$ band at the LSMO interface. In the right panel we illustrate the $\Delta\left(\dfrac{\mathrm{d}I}{\mathrm{d}V}\right)$ and the easy magnetization axis of only the $C_{2[110]}$ component corresponding to $V > -0.2$ V (top) and $V < -0.2$ V (bottom). All the data were measured at 4 K.

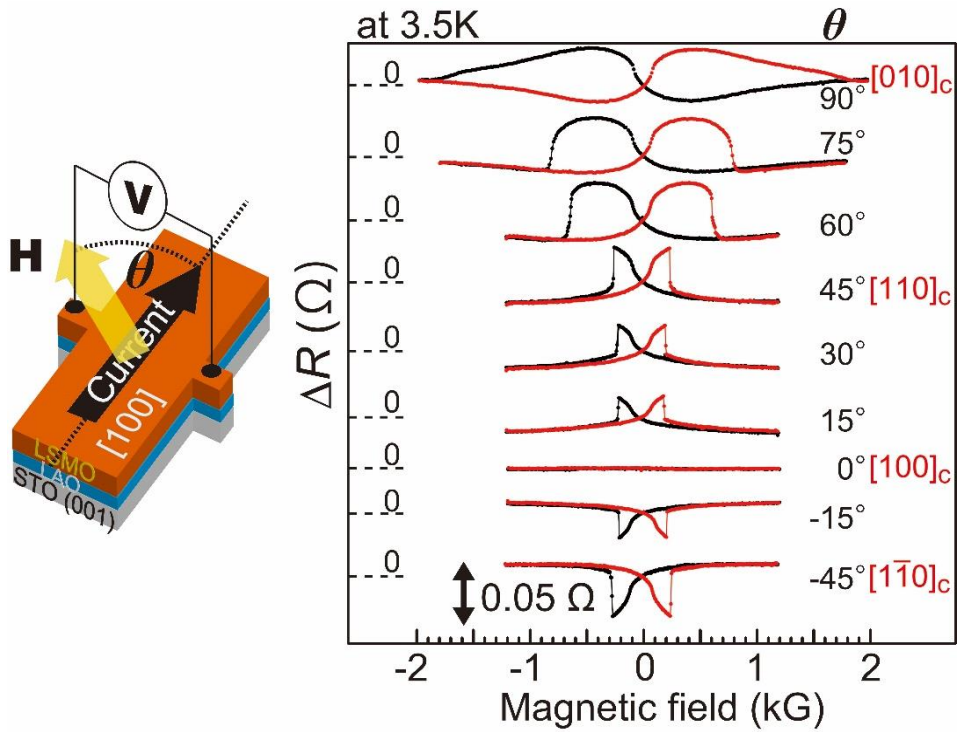

**Figure 3**. **Magnetic anisotropy probed by planar Hall resistance.** (Left panel) Schematic illustration of the Hall bar with a size of $50 \times 200$ μm$^2$ formed along the [100]$_c$ direction using a reference sample of LSMO (40 u.c.)/ LAO (4 u.c.)/ non-doped STO (001) substrate. (Right panel) Planar Hall resistance (PHR) $\Delta R$ of the reference sample with respect to the one at the zero magnetic field measured under various in-plane magnetic field **H** directions. The angle between **H** and the [100]$_c$ direction is denoted as $\theta$.

.

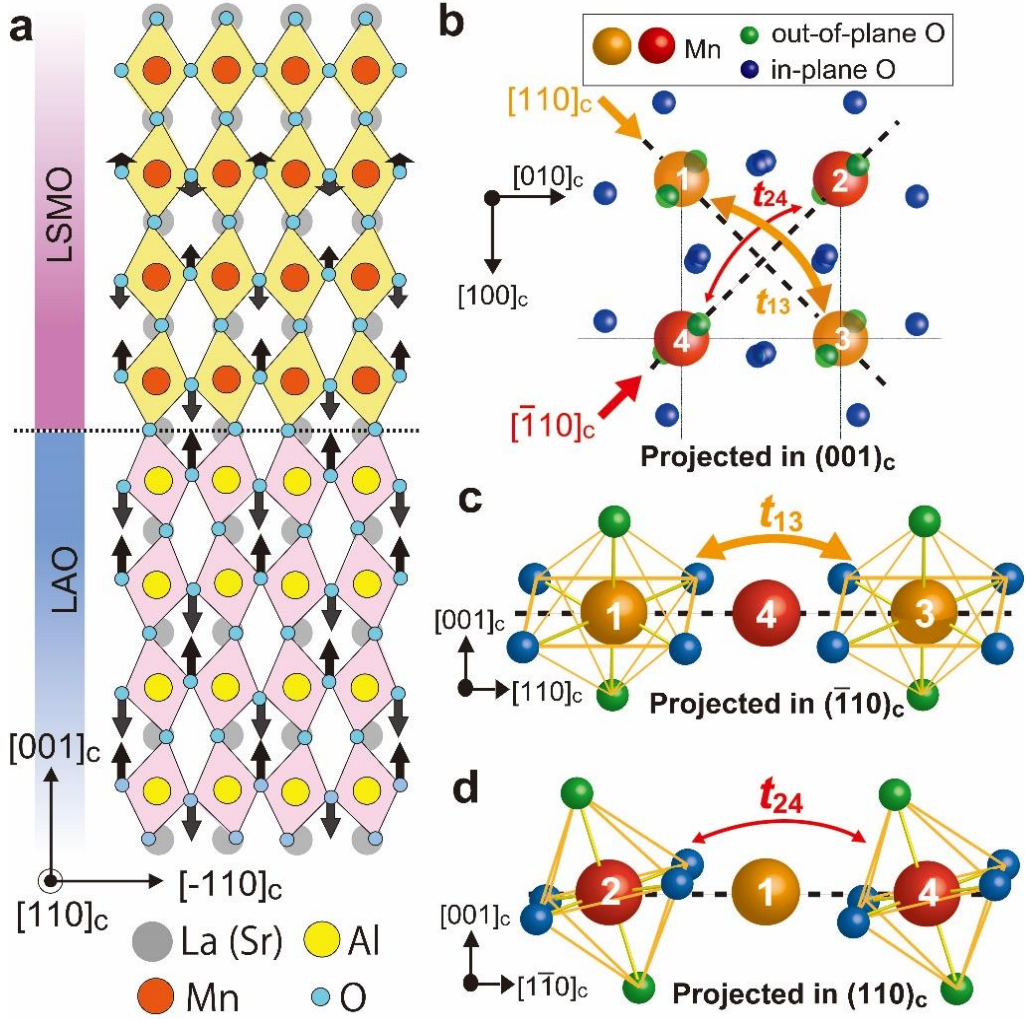

**Figure 4. Origin of the two-fold symmetry magnetic anisotropy along the** $[110]_c$ **direction.** (**a**) Illustration of the crystal structure at the LSMO/LAO interface when looked at from the $[110]_c$ direction. In LAO, adjacent corner-sharing oxygen octahedra rotate around the $[11\bar{1}]_c$ axis in the opposite directions. This lattice distortion is transferred to the first 3–4 u.c. layers of LSMO. (**b–d**). Illustration of four adjacent MnO$_6$ octahedra in a $(001)_c$ plane of LSMO near the LSMO/LAO interface, when projected in the $(001)_c$ (**b**), $(\bar{1}10)_c$ (**c**), and $(110)_c$ (**d**) planes. Due to the OOR around the $[11\bar{1}]_c$ axis, the crystal symmetry between the $[\bar{1}10]_c$ and $[110]_c$ directions is broken. Here, the in-plane and out-of-plane oxygen atoms are drawn in blue and green, respectively. The orange and red spheres represent Mn atoms located along the $[110]_c$ and $[\bar{1}10]_c$ directions, respectively. The rotation angle is largely exaggerated.